\documentclass[prl,amssymb,amsmath,twocolumn,aps,showpacs,superscriptaddress]{revtex4}
\usepackage{amsmath}
\usepackage{graphicx}

\usepackage[dvips]{epsfig}
\usepackage[usenames]{color}
\begin{document}
\title{Diffusion of Interacting Particles in Discrete Geometries}
\author{T. Becker}
\email{thijsbecker@gmail.com}
\affiliation{Hasselt University, B-3590 Diepenbeek, Belgium}
\author{K. Nelissen}
\affiliation{Departement Fysica, Universiteit Antwerpen, Groenenborgerlaan 171, B-2020 Antwerpen, Belgium}
\affiliation{Hasselt University, B-3590 Diepenbeek, Belgium}
\author{B. Cleuren}
\affiliation{Hasselt University, B-3590 Diepenbeek, Belgium}
\author{B. Partoens}
\affiliation{Departement Fysica, Universiteit Antwerpen, Groenenborgerlaan 171, B-2020 Antwerpen, Belgium}
\author{C. Van den Broeck}
\affiliation{Hasselt University, B-3590 Diepenbeek, Belgium}

\date{\today}

\begin{abstract}
We evaluate the self-diffusion and transport diffusion of interacting particles in a discrete geometry consisting of a linear chain of cavities, with
interactions within a cavity described by a free-energy function. Exact analytical expressions are obtained in the absence of correlations, showing that the self-diffusion can exceed the transport diffusion if the free-energy function is concave.
The effect of correlations is elucidated by comparison with numerical results. Quantitative agreement is obtained with recent experimental data for diffusion in a nanoporous zeolitic imidazolate framework material,  ZIF-8.
\end{abstract}
\pacs{05.40.Jc, 02.50.--r, 05.60.Cd, 66.30.Pa}
\maketitle

The equality of inertial and gravitational mass played a crucial role in Einstein's discovery of general relativity. Similarly, Einstein's work on Brownian motion is based on the identity of the transport- and self-diffusion coefficients for noninteracting particles \cite{EinsteinBrown}, leading eventually through Perrin's experiments \cite{Perrin} to the vindication of the atomic hypothesis.  In general, however, diffusion of interacting particles is described by two different coefficients.  The transport-diffusion coefficient $D_t$ quantifies the particle flux $j$ appearing in response to a concentration gradient $dc/dx$:
\begin{equation}\label{transdiff}
j = -D_t  \frac{dc}{dx}.
\end{equation}
The self-diffusion coefficient $D_s$ describes the mean squared displacement of a single particle in a suspension of identical particles at equilibrium: $\langle x^2(t) \rangle \propto D_s t$. An alternative way for measuring this coefficient is by labeling, in this system at equilibrium, a subset of these particles (denoted by $*$) in a way to create a concentration gradient $dc^*/dx$ of labeled particles under overall equilibrium conditions. The resulting flux $j^*$ of  these particles  reads:
\begin{equation}\label{selfdiff2}
j^* = - D_s \frac{dc^*}{dx}.
\end{equation}

Both forms of diffusion have been studied in a wide variety of physical contexts, including continuum \cite{tough1986stochastic,anderson1976diffusion,van1981effect,den1985exact,zwanzig1992diffusion,kwinten2012pre85,kwinten2012epl,kwinten2012pre85bis} and lattice \cite{gomer1990diffusion,ala2002collective, reederlich} models. Exact analytical results for the diffusion coefficient of interacting particles are however typically limited to a perturbation expansion, for example in the density of the particles. The effect of correlations is notoriously difficult to evaluate in continuum models, especially when hydrodynamic interactions come into play, while they can play a dominant role, for example, in lattice models with particle exclusion constraints.

\begin{figure}
\centering
\includegraphics[width=1.0\columnwidth]{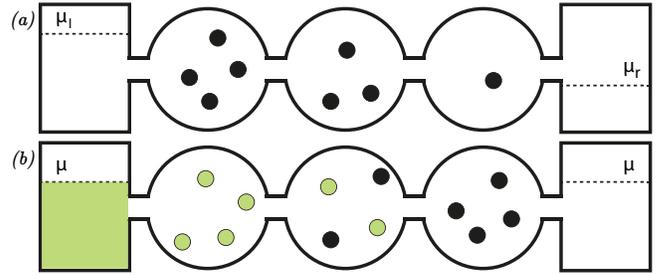}
\caption{(Color online). The model: particles enter cavities via particle reservoirs at certain chemical potential. Particles jump between different cavities through narrow passages. $(a)$ Transport diffusion: a concentration gradient shows a current. $(b)$ Self-diffusion: a concentration gradient of labeled particles is introduced under overall equilibrium conditions.}
\label{fig::selftransdiff}
\end{figure}

In this Letter, we introduce a physically relevant model, for which exact analytical results can be obtained at all values of the concentration and for any interaction. It describes the diffusive hopping of interacting particles in a compartmentalized system, see Fig.\ \ref{fig::selftransdiff} for a schematic representation. It is assumed that the relaxation inside each cavity is fast enough to establish a local equilibrium, described by a free-energy function characterizing the confinement and interaction of the particles.
This model describes diffusion in confined geometries \cite{burada2009}. 
Of particular interest are microporous materials \cite{porousapp,barton1999tailored}, which are widely used in industry, e.g.\ as catalysts in petrochemical industry and as water softeners.
Because of their high thermal and chemical stability \cite{PNASZIF8} and potential applications including carbon dioxide capture and storage \cite{scienceZIF2008} and gas separation \cite{bux2010}, zeolitic imidazolate frameworks (ZIFs) have received considerable interest.  
As illustration, we compare our predictions with experimental results \cite{chmelik10} of diffusion of methanol in ZIF-8.
At variance with previous experiments \cite{prljobic,prlHeinke,prlSalles,angewRosenbach,angewTzoulaki,bookdtds}, it was found that the self-diffusion could exceed the transport diffusion, a result confirmed by molecular dynamics (MD) simulations \cite{cluster1,cluster2,cluster3}. We corroborate the observation that this phenomenon is due to clustering of the particles, and provide an analytical argument and a simple interpretation for the inversion. Our model in fact allows us to reproduce, in a quantitative way, the loading dependence of the self- and transport diffusion for different interactions. Finally, we mention that our model also serves an educational purpose, as the distinction between the transport- and self-diffusion coefficients and the role and contribution of the correlations therein, can be identified explicitly. 

The model consists of a one-dimensional array of pairwise connected cavities, with particles entering in the outer left and right cavities from reservoirs at chemical potentials $\mu_l$ and $\mu_r$ respectively  (see Fig.\ \ref{fig::selftransdiff}). The entire system is at temperature $T$. Particles stochastically jump between cavities by moving through narrow passages. These transitions occur on a slow time scale compared to the relaxation time inside each cavity, ensuring that an equilibrium distribution is effectively maintained in each cavity. It is described by a free energy $F(n) = U(n) - T S(n)$, with $n$ the number of particles in the cavity, $U(n)$ the energy and $S(n)$ the entropy. Furthermore, the dynamics is Markovian with a transition rate which has to reproduce the thermal equilibrium state $p^{\mathrm{eq}}$, when a cavity is connected to a single reservoir at chemical potential $\mu$:
\begin{equation}\label{eqdistro}
p^{\mathrm{eq}}_n(\mu) = ( \mathcal{Z})^{-1}e^{- \beta \left[ F(n)-\mu n \right]},
\end{equation}
with $\beta=(k_BT)^{-1}$ and \( {\mathcal{Z}}^{-1} \) the normalization constant.

We first derive an exact expression for $D_t$ and $D_s$, in a limiting situation where correlations between particle numbers in different cavities are absent (see also supplementary material \cite{supplmat}).
Consider a system consisting of three cavities, with $n_l$, $n$ and $n_r$ specifying the number of particles inside the left, middle and right cavity, respectively. In the  limit in which the exchange rates with the middle cavity are small compared to the exchange rates with the reservoirs, the left and right cavity are effectively decorrelated from the middle cavity, and are characterized by the equilibrium probability distribution $p^{\mathrm{eq}}_{n_l}(\mu_l)$  and $p^{\mathrm{eq}}_{n_r}(\mu_r)$, respectively. This setup allows us to obtain exact analytical results at arbitrary particle density. The probability distribution $p_n$ for the middle cavity obeys the following master equation:
\begin{equation}\label{mastereq}
\dot{p}_n = k^+_{n-1} p_{n-1} + k^-_{n+1} p_{n+1} - \left( k^+_n + k^-_n \right) p_n,
\end{equation}
with $k^+_n$ and $k^-_n$ the rates to add or remove a particle from the middle cavity containing $n$ particles: 
\begin{align} 
k^+_n &= \sum_{n_l} p^{\mathrm{eq}}_{n_l}(\mu_l) k_{n_l n} + \sum_{n_r} p^{\mathrm{eq}}_{n_r}(\mu_r) k_{n_r n} \label{kp} \\
k^-_n &= \sum_{n_l} p^{\mathrm{eq}}_{n_l}(\mu_l) k_{n n_l} + \sum_{n_r} p^{\mathrm{eq}}_{n_r}(\mu_r) k_{n n_r}.  \label{km}
\end{align}
The rate $k_{n m}$ denotes the probability per unit time for a particle to jump from a cavity containing $n$ particles to a neighboring cavity containing $m$ particles. At this stage we do not need to specify its explicit form, but we request that it obeys detailed balance:
\begin{equation}\label{localdetbal}
k_{n m} / k_{m+1,n-1} = e^{- \beta \left[ F(m+1) + F(n-1) - F(n) - F(m)  \right]}.
\end{equation}
The particle flux and concentration difference between the left and middle cavity read:
\begin{align}
j(\mu_l,\mu_r) &= \sum_{n,n_l} \left( k_{n_l n} - k_{n n_l}  \right) p^{\mathrm{eq}}_{n_l} (\mu_l) p_n, \label{jreal} \\
dc(\mu_l,\mu_r) &= \left( 1 / \lambda \right) \sum_{n,n_l}(n-n_l) p^{\mathrm{eq}}_{n_l} (\mu_l) p_n  \label{dcreal},
\end{align}
where $\lambda$ is the center-to-center distance between cavities.
The transport diffusion $D_t$, quantifying the linear response of $j$ with respect to $dc$,  is found from the ratio $-j/(dc/ \lambda)$ in the limit $\delta=(\mu_l - \mu_r)/2 \rightarrow 0$. Introducing the average chemical potential $\mu = (\mu_l+\mu_r)/2$, 
one finds for Eqs.\ \eqref{kp} and \eqref{km} up to linear order in $\delta$:
\begin{equation}\label{ratefirstorder}
k^+_n = 2 \sum_{m} p^{\mathrm{eq}}_{m}(\mu) k_{m n}, \quad k^-_n = 2 \sum_{m} p^{\mathrm{eq}}_{m}(\mu) k_{n m}. 
\end{equation}
One concludes from Eq.\ \eqref{mastereq} that at this order in $\delta$, the steady state solution of the master equation is given by $p_n = p^{\mathrm{eq}}_n(\mu)$. The corresponding current and concentration difference are obtained from the expansion of Eqs.\ \eqref{jreal} and \eqref{dcreal} to first order in $\delta$, resulting in
\begin{equation}\label{dtfinal}
D_t(\mu) = \frac{\lambda^2 \sum_{n,m} p_n^{\mathrm{eq}}(\mu) p_{m}^{\mathrm{eq}}(\mu) k_{n m} }{  \langle n^2 \rangle - \langle n \rangle^2  } \equiv   \frac{ \lambda^2 \langle k \rangle}{  \langle n^2 \rangle - \langle n \rangle^2 },
\end{equation}
where $\langle \cdot \rangle$ denotes the average over $p^{\mathrm{eq}}(\mu)$.

We next turn to the self-diffusion, using the labeling procedure discussed in the introduction. Since the final expression for $D_s$ does not depend on the labeling percentages, we consider a simple case: all particles in the left reservoir are labeled, those in the right reservoir remain unlabeled. As a result, all particles in the left and none in the right cavity are labeled.
The state of the middle cavity is now described by two numbers, $n$ (total number of particles) and $n^*$, the number of labeled particles. The corresponding steady state probability distribution $p_{n,n^*}$ is:
\begin{equation}
p_{n,n^*} = p^{\mathrm{eq}}_n(\mu)\frac{n!}{n^*!(n-n^*)!}\frac{1}{2^n}.
\end{equation}
The flux of labeled particles and concentration difference between the left and middle reservoir read:
\begin{align}
j^* &= \sum_{n_l,n,n^*} \left( k_{n_l n}- k_{n n_l} \frac{n^*}{n} \right) p_{n,n^*} p^{\mathrm{eq}}_{n_l}(\mu) = \frac{ \langle k  \rangle}{2} \nonumber\\
dc^* &= \left( 1 / \lambda \right) \sum_{n_l,n,n^*} (n^* - n_l ) p_{n,n^*} p^{\mathrm{eq}}_{n_l}(\mu)= -\langle n \rangle/2.
\end{align}
Hence, the self-diffusion $D_s =- j^*/(dc^*/\lambda)$ reads
\begin{equation}\label{dsfinal}
D_s(\mu) = \lambda^2 \langle k  \rangle/\langle n \rangle.
\end{equation}
Equations \eqref{dtfinal} and \eqref{dsfinal} constitute the main analytical results in this Letter. They are valid at all values of the concentration and can be calculated for any interaction.
From Eqs.\ \eqref{dtfinal} and \eqref{dsfinal}, one finds for the ratio of $D_t$ and $D_s$:
\begin{equation}\label{ratioselftrans}
\frac{D_t(\mu)}{D_s(\mu)} = \frac{ \langle n \rangle}{ \langle n^2 \rangle - \langle n \rangle^2} = \frac{\langle n \rangle}{\text{Var}(n)} = \Gamma(\mu) ,
\end{equation}
where $\Gamma(\mu)$, the so-called thermodynamic factor, is an equilibrium property. 
Equation \eqref{ratioselftrans} can be derived by a general argument, when correlations are ignored \cite{gomer1990diffusion,ala2002collective}. We note that Eqs.\ \eqref{dtfinal} and \eqref{dsfinal} remain valid for any number of cavities between the left and right cavities, provided correlations in particle number are ignored  \cite{preparation}.
 
We now comment on the effect of interaction, and in particular of the shape of the free energy, on the thermodynamic factor. In the absence of interactions, the free energy is that of an ideal gas  $\beta F^{\mathrm{id}}(n) = \ln(n!) + c n$, with $c$ a constant \cite{kardarint}.
The corresponding distribution $p^{\mathrm{eq}}_n(\mu)$ is Poissonian for which $\langle n \rangle = \text{Var}(n)$ and hence $\Gamma = 1$. One recovers the ``Einstein''  result  that $D_s = D_t$ for noninteracting particles.
Note that adding an arbitrary linear term $\propto n$ to $F(n)$ corresponds to a rescaling of the chemical potential, see Eq.\ \eqref{eqdistro}. Hence, a linear term in $F(n)$ does not influence the statistics at a given loading $\langle n \rangle$. 
We now show that for deviations of the free energy  from the ideal gas value, $f(n) = F(n) - F^{\mathrm{id}}(n)$, 
the ratio of $D_t$ and $D_s$ is determined by the convexity versus concavity of $f(n)$, where we will call $f(n)$ the interaction free energy. 
We fix the loading $\langle n \rangle$ and consider two neighboring cavities containing, respectively, $n_1$ and $n_2$ $(> n_1)$ particles. A particle now jumps so that the new state becomes $n_1-1$ and $n_2+1$. Such an event increases the local density inhomogeneity. When $f(n)$ is convex ($f''(n)>0$), the interaction free energy is larger in the new state: $f(n_1) + f(n_2)<f(n_1 - 1) + f(n_2+1)$.  Therefore its probability is small compared to the situation with no interactions. $\text{Var}(n)$ decreases since particle numbers different from the average loading become less probable. Hence when $f(n)$ is convex, $\text{Var}(n)<\langle n \rangle$, $\Gamma > 1$ and $D_t > D_s$. When $f(n)$ is concave ($f''(n)<0$) the opposite happens. The interaction free energy of the new state is smaller and both its probability and $\text{Var}(n)$ increase, leading to  $\Gamma < 1$ and $D_t < D_s$.

A few additional remarks are in order. First,  a cavity can typically contain a limited  number of particles $n\leq n_{\mathrm{max}}$. This corresponds to  $f(n) = \infty$ for all $n > n_{\mathrm{max}}$, i.e., $f(n)$ is ``infinitely convex'' at $n_{\mathrm{max}}$.  We conclude from the above argument that a concave section is a necessary, but not sufficient, condition for having Var($n$) $> \langle n \rangle$, i.e.,  for $D_s$ to exceed $D_t$. 
Second, one can give an intuitive explanation as to why a concave $f(n)$ promotes $D_s > D_t$.
$D_t$ is measured by a flux $j$.
If $f(n)$ is concave, particles tend to cluster, which will mostly happen in cavities that are already high in particle number. This causes the particles to be ``pulled back'' towards the region of higher concentration. The net effect is a force in the direction of higher concentration, lowering the particle flux. 
$D_s$ is measured by a flux of labeled particles $j^*$.
Since the system is in equilibrium there is no concentration gradient. As a result, there is no preferential direction for clustering, and there will be no force counteracting the current of labeled particles.
Finally, the experimental finding of $D_s$ exceeding $D_t$ \cite{chmelik10} was explained on the basis of MD simulations \cite{cluster1,cluster2,cluster3} as due to clustering of particles. Our model corroborates this conclusion but in addition provides a simple physical interpretation and an analytical argument.

\begin{figure}
\centering
\includegraphics[width=1.0\columnwidth]{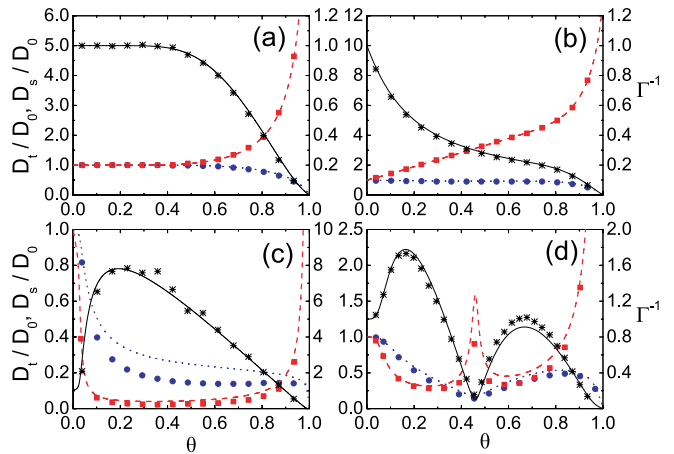}
\caption{(Color online). $D_s/D_0$, $D_t/D_0$ and $\Gamma^{-1}$ as a function of loading $\theta=\langle n \rangle/n_{\mathrm{max}}$, $n_{\mathrm{max}} = 13$; for (a) $\beta f(n)=0$, (b) $\beta f(n) = 0.2 n^2$, (c) $\beta f(n) = - 0.2 n^2$ and (d) $\beta f(n)$ that is subsequently concave, convex and again concave. 
The red dashed lines (analytical solution) and squares (simulations) show the transport diffusion and the blue dotted lines  (analytical solution) and full circles (simulations) the self-diffusion (values on lhs axis).
The analytical $\Gamma^{-1}$ (black full lines) are compared with the ratio of $D_s$ and $D_t$ (black stars) from the simulations (values on rhs axis).}
\label{fig::simdiff}
\end{figure}

For systems containing an arbitrary number of cavities, one has to take into account correlation effects. Finding an exact solution becomes difficult. Instead, we have performed kinetic Monte Carlo simulations (see supplementary material \cite{supplmat}). 
Our choice of rates is:
\begin{equation}
k_{n m}= \nu n e^{ - \frac{\beta}{2} \left[ f(n-1) + f(m+1) - f(n) - f(m)  \right]}.
\end{equation}
The factor $\nu$ determines the time scale. In the limit of infinite dilution, both $D_t$ and $D_s$ are equal to $\nu \lambda^2 \equiv D_0$. For an ideal gas $f(n) = 0$, $k_{n m} = \nu n$, i.e., the rates satisfy the law of mass action \cite{vankampen1981}. The simulations presented here are for 15 pairwise connected cavities,  with $n_{\mathrm{max}} = 13$.
$D_s/D_0$, $D_t/D_0$ and $\Gamma^{-1}$  are plotted in Fig.\ \ref{fig::simdiff} for different types of free energies, as a function of the loading $\theta=\langle n \rangle / n_{\mathrm{max}}$. Both the simulation data and the analytical curves Eqs.\ \eqref{dtfinal} and \eqref{dsfinal} are shown. The stars in the figures correspond to the ratio between the self- and transport diffusion obtained from simulations. 
Since correlations are included in the simulations but absent in the analytical result, the difference of the two curves is a measure of the effect of correlations on the diffusion.
Figure \ref{fig::simdiff}(a) shows the diffusion for noninteracting particles $\beta f(n) = 0$, with confinement (presence of $n_{\mathrm{max}}$). At low and medium loadings the particles are not influenced by the confinement; $\Gamma = 1$ and $D_s = D_t$. At high loading, the confinement comes into play: $\Gamma^{-1}$ decreases, $D_t$ rises and $D_s$ lowers. The effect of correlations is negligible: the simulation data and analytical results coincide almost  perfectly.
Figure \ref{fig::simdiff}(b) shows the diffusion in the case of a convex free energy $\beta f(n) = 0.2 n^2$. $\Gamma^{-1}$ is lower than one, and $D_t$ is always larger than $D_s$. Correlations have a negligible influence.
Fig.\ \ref{fig::simdiff}(c) shows the diffusion for a concave free energy $\beta f(n) = - 0.2 n^2$. As expected,  $D_s > D_t$ for low to moderate loading. At moderate and high loading the ``convexity effect'' of confinement takes over: $\Gamma^{-1}$ decreases and eventually becomes smaller than one with $D_t > D_s$. 
This curve should be compared with Figs.\ $3 (a),(c)$ in \cite{chmelik10}.
Noteworthy is the fact that the transport diffusion shows a minimum when the thermodynamic factor is around its maximum. This feature is in agreement with experimental observations  \cite{chmelik10,chmelik09,ACSsalles} and with MD simulations \cite{DTminKrishna}. It is now easily understood: when $\Gamma^{-1}$ is at its highest, the tendency to cluster is maximal, therefore the force opposing the current is also at its strongest. Turning to the effect of correlations, we note that they are quite strong: both $D_t$ and $D_s$ are significantly lower than the analytical results. The effect is the largest for self-diffusion. 
Nevertheless, the ratio of $D_t$ and $D_s$ is still very close to $\Gamma$, again in agreement with what is observed in experiments \cite{chmelik10} and MD simulations in similar systems \cite{krishnacorrelation}.
Fig.\ \ref{fig::simdiff}(d) shows the diffusion for a free energy that is first concave, then convex and then concave (see supplementary material \cite{supplmat} for the exact form). For the first concave part the self-diffusion exceeds the transport diffusion. For the second concave part this is no longer the case, due to the confinement and the influence of the convex part in the middle. This is an illustration of how concavity is necessary but not sufficient for $D_s > D_t$. $D_t$ shows a (local) maximum in the convex part, whereas $D_s$ shows a (local) minimum.
Correlations have noticeable effect, and are now more important for $D_t$ than for $D_s$. 
Notice that in all cases correlations lower the diffusion coefficients.

\begin{figure}
\centering
\includegraphics[width=1.0\columnwidth]{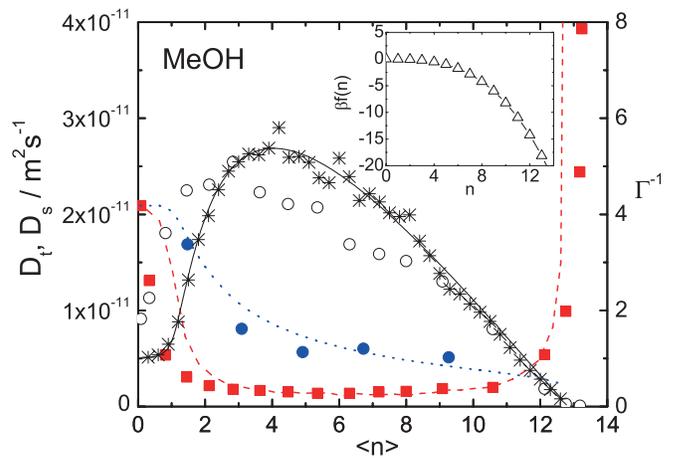}
\caption{(Color online). Comparison of experimental data of methanol in ZIF-8 \cite{communicationChmelik} with simulations from our model.
The experimental self-diffusion and transport diffusion are represented by the blue full circles and red squares, respectively, with the corresponding results from simulations given by the blue dotted line and the red dashed line (values on lhs axis).
The experimental (black open circles) and analytical (full black line) $\Gamma^{-1}$ are compared with the ratio of self- and transport diffusion (black stars) taken from the simulations (values on rhs axis). The inset shows $\beta f(n)$.}
\label{fig::expdiff}
\end{figure}

Motivated by the qualitative agreement with experiments, we have tried to reproduce the experimental results from \cite{chmelik10} quantitatively. Inspired by the form of the energy function for Lennard-Jones crystals \cite{energyLJ}, we  take $\beta f(n) = a n^2 + b n^3$ for $n \leq n_{\mathrm{max}}$, with $n_{\mathrm{max}} = 13$ taken from the experimental data \cite{communicationChmelik}.
The parameters $a$ and $b$ are determined by fitting the thermodynamic factor (Eq.\ \eqref{ratioselftrans}) with the experimental data, resulting in $\beta f(n) = 0.000642 n^2 - 0.0083 n^3$.
The parameters $\nu$ and $\lambda$ only appear in the combination $\nu \lambda^2$, which follows directly from the experimental value of $D_t$ at very low loading. In Fig.\ \ref{fig::expdiff} we compare the obtained simulation results for $D_s$ and $D_t$ with experimental data of methanol in ZIF-8 \cite{communicationChmelik}. Quantitative agreement is found for both $D_s$ and $D_t$ at all values of the loading. This is remarkable since $a$ and $b$ are determined from the equilibrium quantity $\Gamma$, and only the experimental value of $D_t$ at very low loading is used in the fit of $\nu \lambda^2$. A similar quantitative agreement is also found for ethanol in ZIF-8 (cf. supplementary material \cite{supplmat}).

To conclude, we have introduced a model describing diffusion of interacting particles in discrete geometries. 
Exact analytical expressions for the self- and transport diffusion are given in the limiting case where correlations are absent, but are otherwise valid at all values of the concentration and for any interaction.
 We showed that the self-diffusion can exceed the transport diffusion when the free-energy function is  concave as a function of the loading, resulting in the clustering of  particles. By comparison with numerical simulations, the effect of the correlations is elucidated. Their influence is found to be  significant for a free energy that is very concave or has several convex and concave sections. Nevertheless the ratio of self- and  transport diffusion is always close to the thermodynamic factor, $D_t/D_s \approx \Gamma$, a result which is exact in the absence of correlations. Finally, we obtained quantitative agreement between numerical simulations of our model and experimental results of diffusion in ZIF-8 from Ref.\ \cite{chmelik10}.

\begin{acknowledgments}
This work was supported by the Flemish Science Foundation (FWO-Vlaanderen).
\end{acknowledgments}

\end{document}


\title{Diffusion of Interacting Particles in Discrete Geometries \\ Supplemental Material}
\author{T. Becker}
\affiliation{\it Hasselt University, B-3590 Diepenbeek, Belgium}
\author{K. Nelissen}
\affiliation{\it Departement Fysica, Universiteit Antwerpen, Groenenborgerlaan 171, B-2020 Antwerpen, Belgium}
\affiliation{\it Hasselt University, B-3590 Diepenbeek, Belgium}
\author{B. Cleuren}
\affiliation{\it Hasselt University, B-3590 Diepenbeek, Belgium}
\author{B. Partoens}
\affiliation{\it Departement Fysica, Universiteit Antwerpen, Groenenborgerlaan 171, B-2020 Antwerpen, Belgium}
\author{C. Van den Broeck}
\affiliation{\it Hasselt University, B-3590 Diepenbeek, Belgium}

\pacs{05.40.Jc, 02.50.--r, 05.60.Cd, 66.30.Pa}


\maketitle

\tableofcontents

\section{Separation of time-scales}

It is shown how to obtain the master equation for a system with time-scale separation, Eq.\ $(4)$ in the Letter. 
The theory and notation from \cite{Esposito12} is used. The system consists of three cavities. The left and right cavity are connected to particle reservoirs. The state is denoted by $(n_l, n ,n_r)$, the number of particles in respectively the left, middle, and right cavity. Following notation from \cite{Esposito12}, ``microstates'' are denoted by $(n_l, n ,n_r)$, and ``mesostates'' are denoted by $n$, the number of particles in the middle cavity. 
By time-scale separation, we mean that the dynamics between different microstates belonging to the same mesostate is much faster than between microstates belonging to different mesostates. In our model, this means that transitions between the reservoirs and the cavities are much faster than transitions between the cavities. 
The probability to find the system in mesostate $n$ equals
\begin{equation}
P_n = \sum_{n_l,n_r} p(n_l,n,n_r).
\end{equation}   
The conditional probability to be in microstate $(n_l,n,n_r)$ being in the mesostate $n$ is given by
\begin{equation}
{\mathbb{P}}_n(n_l,n_r) = p(n_l,n,n_r)/ P_n.
\end{equation}
Due to the time-scale separation, the ${\mathbb{P}}_n(n_l,n_r)$ evolve much faster than the mesostate probabilities $P_n$. The ${\mathbb{P}}_n(n_l,n_r)$'s obey an almost isolated dynamics inside the mesostate $n$, eventually relaxing to the stationary distribution ${\mathbb{P}}^{\mathrm{st}}_n(n_l,n_r)$:
\begin{equation}\label{statcondprob}
\sum_{n'_l,n'_r} W_{(n_l,n,n_r),(n'_l,n,n'_r)}  {\mathbb{P}}^{\mathrm{st}}_n(n'_l,n'_r) = 0,
\end{equation}
where $W_{(n_l,n,n_r),(n'_l,n,n'_r)}$ is the rate to jump from $(n'_l,n,n'_r)$ to $(n_l,n,n_r)$. This rate can be separated into a sum of rates due to the left and right reservoir
\begin{align}
&\sum_{n'_l,n'_r} W_{(n_l,n,n_r),(n'_l,n,n'_r)}  {\mathbb{P}}^{\mathrm{st}}_n(n'_l,n'_r) \\
= &\sum_{n'_l,n'_r} \left( W^{(l)}_{(n_l,n,n_r),(n'_l,n,n_r)} + W^{(r)}_{(n_l,n,n_r),(n_l,n,n'_r)}  \right)  {\mathbb{P}}^{\mathrm{st}}_n(n'_l,n'_r), \label{leftright2}
\end{align}
where transitions with the left (right) reservoir only change $n_ l$ $(n_r)$. Because particles only interact within the same cavity, transition rates of the reservoirs only depend on the particle number of the cavity they are connected to. Eq.\ \eqref{leftright2} can be rewritten as
\begin{align}\label{leftright3}
\sum_{n'_l,n'_r} \left( W^{(l)}_{n_l,n'_l} + W^{(r)}_{n_r,n'_r} \right) {\mathbb{P}}^{\mathrm{st}}_n(n'_l,n'_r).
\end{align}
These rates are independent of $n$, so ${\mathbb{P}}^{\mathrm{st}}_n(n_l,n_r)$ does not depend on $n$, allowing us to write ${\mathbb{P}}^{\mathrm{st}}(n_l,n_r)$. Moreover, $W^{(l)}_{n_l,n'_l}$ and $W^{(r)}_{n_r,n'_r}$ are not influenced by each other. The probabilities to have $n_l$ or $n_r$ particles are therefore uncorrelated, and we can write ${\mathbb{P}}^{\mathrm{st}}(n'_l,n'_r) = {\mathbb{P}}^{\mathrm{st}}(n'_l) {\mathbb{P}}^{\mathrm{st}}(n'_r)$. Eq.\ \eqref{leftright3} can be written as
\begin{align}
&\sum_{n'_l,n'_r} \left( W^{(l)}_{n_l,n'_l}  {\mathbb{P}}^{\mathrm{st}}(n'_l) {\mathbb{P}}^{\mathrm{st}}(n'_r) + W^{(r)}_{n_r,n'_r}  {\mathbb{P}}^{\mathrm{st}}(n'_l) {\mathbb{P}}^{\mathrm{st}}(n'_r)  \right) \\
= &\sum_{n'_l}  W^{(l)}_{n_l,n'_l}  {\mathbb{P}}^{\mathrm{st}}(n'_l)   + \sum_{n'_r}  W^{(r)}_{n_r,n'_r}  {\mathbb{P}}^{\mathrm{st}}(n'_r) = 0.
\end{align}
The rates of the left $(l)$ and right $(r)$ reservoir obey local detailed balance
\begin{equation}
\frac{W^{(l,r)}_{i,i+1}}{W^{(l,r)}_{i+1,i}} = \exp \left( - \frac{F(i) - \mu_{(l,r)} i - F(i+1) +  \mu_{(l,r)} (i+1)  }{k_b T}   \right).
\end{equation}
${\mathbb{P}}^{\mathrm{st}}(n_l)$ and ${\mathbb{P}}^{\mathrm{st}}(n_r)$ are therefore equal to the equilibrium probability distributions $p^{\mathrm{eq}}_{n_l}(\mu_{l})$ and $p^{\mathrm{eq}}_{n_r}(\mu_{r})$. The end result reads
\begin{equation}
{\mathbb{P}}^{\mathrm{st}}_n(n_l,n_r) = p^{\mathrm{eq}}_{n_l}(\mu_l) p^{\mathrm{eq}}_{n_r}(\mu_r).
\end{equation}
Using first-order perturbation theory (see Appendix A of \cite{Esposito12}), one can show that the master equation of $P_n$ is given by
\begin{equation}\label{timescalefinalsol}
\dot{P}_n = V^{\mathrm{st}}_{n,n-1} P_{n-1} + V^{\mathrm{st}}_{n,n+1} P_{n+1} - (V^{\mathrm{st}}_{n-1,n} + V^{\mathrm{st}}_{n+1,n}) P_{n},
\end{equation}
with
\begin{align}
V^{\mathrm{st}}_{n+1,n} &=  \sum_{n_l,n_r,n'_l,n'_r} W_{(n_l,n+1,n_r),(n'_l,n,n'_r)}   {\mathbb{P}}^{\mathrm{st}}_n(n'_l,n'_r)   \\
&= \sum_{n_l,n_r,n'_l,n'_r} W_{(n_l,n+1,n_r),(n'_l,n,n'_r)}  p^{\mathrm{eq}}_{n'_l}(\mu_l) p^{\mathrm{eq}}_{n'_r}(\mu_r) \\
&= \sum_{n'_l, n'_r} (k_{n'_l n} + k_{n'_r n})  p^{\mathrm{eq}}_{n'_l}(\mu_l) p^{\mathrm{eq}}_{n'_r}(\mu_r) \\
&=  \sum_{n'_l} p^{\mathrm{eq}}_{n'_l}(\mu_l) k_{n'_l n} + \sum_{n'_r} p^{\mathrm{eq}}_{n'_r}(\mu_r) k_{n'_r n},
\end{align}
which is the equation for $k^+_n$ in the Letter. $k^-_n$ is found similarly. One then finds that Eq.\ \eqref{timescalefinalsol} equals Eq.\ $(4)$ in the Letter.

\section{Self-diffusion: arbitrary percentages of labeled particles}

Again consider three cavities with the separation of time-scales discussed in the previous section. Suppose $\alpha$ percent of the particles is labeled in the left cavity, and $\beta$ percent in the right cavity. Due to the time-scale separation, these percentages are constant. The system is in equilibrium at chemical potential $\mu$. The stationary probability $p_{n,n^*}$ to find $n$ particles of which $n^*$ are labeled in the middle cavity equals
\begin{equation}\label{sollabel}
p_{n,n^*} =  p^{\mathrm{eq}}_n(\mu) \binom{n}{n^*} \left( \frac{\alpha + \beta}{2} \right)^{n^*} \left(1-Ê \frac{\alpha + \beta}{2} \right)^{n-n^*}  \equiv  p^{\mathrm{eq}}_n(\mu)  B_{n,(\alpha + \beta)/2}(n^*).
\end{equation}
$B_{n,(\alpha + \beta)/2}$ is the binomial distribution with parameters $n$ and $(\alpha + \beta)/2$. The average of $B_{n,\alpha}(n^*)$ equals $\alpha n$.
One can understand that this is the correct distribution from a combinatorial argument. The probability to have $n$ particles is given by the equilibrium distribution $p^{\mathrm{eq}}_n(\mu)$; the labeling of the particles has no influence on this result. What is the probability to have $n^*$ labeled particles if there are $n$ particles in the cavity? A particle that enters the middle cavity has equal probability to have come from the left or right cavity, since the system is in equilibrium. The probability that a particle entering from the left is labeled equals $\alpha$; when it enters from the right this probability is $\beta$. The total probability that a particle entering the middle cavity is labeled is therefore $(\alpha + \beta)/2$. The probability to have $n^*$ labeled particles when there are $n$ particles in the middle cavity equals
\begin{equation}\label{binomial}
\binom{n}{n^*} \left( \frac{\alpha + \beta}{2} \right)^{n^*} \left(1-Ê \frac{\alpha + \beta}{2} \right)^{n-n^*}  \equiv B_{n,(\alpha + \beta)/2}(n^*).
\end{equation}
This is the binomial distribution $B_{n,(\alpha + \beta)/2}(n^*)$. It should be interpreted as to probability to win $n^*$ times out of $n$ tries, when the probability to win equals $(\alpha + \beta)/2$. $p_{n,n^*}$ is found by multiplying Eq.\ \eqref{binomial} with $p^{\mathrm{eq}}_n(\mu)$. For $\alpha = 1$ and $\beta = 0$ one recovers the result from the Letter.

To verify this is the correct solution, one can solve the master equation for $p_{n,n^*}$. The rate for an unlabeled particle to enter the middle cavity equals $k^+_n \left( 1 -  (\alpha + \beta)/2 \right) $; a labeled particle enters with rate $k^{+}_n (\alpha + \beta)/2 $. The rates to leave the middle cavity depend on the state $(n,n^*)$: An unlabeled particle leaves the middle cavity at rate $k^-_n \left( (n-n^*)/n  \right)$; a labeled particle leaves the middle cavity with rate $k^-_n \left( n^*/n  \right)$. The master equation reads
\begin{align}
\dot{p}_{n,n^*} = &\left( 1 - \frac{ \alpha + \beta}{2} \right)  k^+_{n-1} p_{n-1,n^*} +  \frac{\alpha + \beta}{2} k^{+}_{n-1}  p_{n-1,n^*-1}  \notag \\
+ &\frac{n+1-n^*}{n+1}   k^-_{n+1} p_{n+1,n^*} + \frac{n^*+1}{n+1} k^-_{n+1} p_{n+1,n^*+1}  \notag \\
- &\left( k^+_n + k^-_n   \right) p_{n,n^*}.
\end{align}
Using the probability distribution Eq.\ \eqref{sollabel} for $p_{n,n^*}$, this equation reduces to
\begin{equation}\label{checkself}
\dot{p}_{n,n^*} = k^+_{n-1} p^{\mathrm{eq}}_{n-1} + k^-_{n+1} p^{\mathrm{eq}}_{n+1} - \left( k^+_n + k^-_n \right) p^{\mathrm{eq}}_n = \dot{p}^{\mathrm{eq}}_{n} = 0.
\end{equation}
One finds
\begin{align}
\lambda dc^* &= \sum_{n_l,n_l^*,n,n^*} (n^*-n_l^*) p^{\mathrm{eq}}_{n_l}(\mu) p^{\mathrm{eq}}_n(\mu) B_{n_l,\alpha}(n_l^*) B_{n,(\alpha + \beta)/2}(n^*) \\
&= \frac{\beta - \alpha}{2} \langle n \rangle,
\end{align}
and
\begin{align}
j^* &= \sum_{n_l,n_l^*,n,n^*} \left( k_{n_l n} \frac{n_l^*}{n_l}  - k_{n n_l} \frac{n^*}{n} \right) p^{\mathrm{eq}}_{n_l}(\mu) p^{\mathrm{eq}}_n(\mu) B_{n_l,\alpha}(n_l^*) B_{n,(\alpha + \beta)/2}(n^*) \\
&= -  \frac{\beta - \alpha}{2} \langle n \rangle.
\end{align}
The end result reads
\begin{equation}
D_s = \lambda^2 \langle k \rangle / \langle n \rangle.
\end{equation}

\section{Simulation details}

\subsection{Putting the system at a certain loading}

On the left and right, the system is connected to particle reservoirs. The rates to jump from/to a reservoir at chemical potential $\mu$ equal
\begin{align}  
k( n \rightarrow n+1 ) &= \nu \exp \left(  - \frac{\beta}{2} \left[ f(n+1)-f(n) \right] + \beta \mu   \right)  \\ 
k( n \rightarrow n-1 ) &= \nu n  \exp \left(  - \frac{ \beta}{2} \left[ f(n-1)- f(n) \right]   \right), \label{rateres2} 
\end{align}
where $n$ is the number of particles in the connected cavity.

The loading $\langle n \rangle (\mu) = \sum_i i p^{\mathrm{eq}}_i(\mu)$ of the system is determined by the chemical potentials of the reservoirs, and can be calculated analytically from $p^{\mathrm{eq}}_n(\mu)$. It is however not possible to calculate the inverse of $\langle n \rangle (\mu)$, i.e., to find $\mu$ corresponding to a given $\langle n \rangle$. We therefore do this numerically. We write a program which takes as input the desired loadings that need to be simulated and gives as output the associated chemical potentials. This list of chemical potentials is used as input for the program. 

In the limit of an infinite number of cavities the self-diffusion and transport diffusion converge to a limiting value. The diffusion curves in the paper are obtained from a line of 15 cavities. For this length, the diffusion coefficients have almost completely converged to the limiting value.

We have also done simulations where the cavities on the boundaries exchange particles with the reservoirs on a much faster timescale than particles are exchanged between the cavities, similar to the theory in the letter. The difference between the diffusion curves for the two situations was negligible. 

\subsection{Measuring the self-diffusion}

The chemical potentials of both reservoirs are taken equal, bringing the whole system at the same loading $\langle n \rangle$. Particles entering from the left reservoir are labeled with a certain percentage, and particles entering from the right reservoir are labeled with a different percentage. The flux of labeled particles and concentration difference of labeled particles are measured between all cavities. These values are averaged, and are used to calculate $D_s$.
We have chosen to label all particles in the left reservoir and none in the right reservoir.
It was checked that the percentages at which the particles are labeled is of negligible influence on the end result.

\subsection{Measuring the transport diffusion}

To measure $D_t$ at a certain loading $\langle n \rangle$, one puts the left and right reservoir at different chemical potentials. One chooses $\mu_l$ that gives a loading $\langle n \rangle + \delta \langle n \rangle$, and  $\mu_r$ that gives a loading $\langle n \rangle - \delta \langle n \rangle$, with $ \delta \langle n \rangle$ small. This assures that one is still in the regime of linear response. (In our simulations $ \delta \langle n \rangle \approx 0.2$ for $n_{max} = 13$.) The particle flux and concentration difference are measured between all cavities. These values are averaged, and are used to calculate $D_t$. The loading is taken as the average loading of all cavities, and will be $\approx \langle n \rangle$.

\subsection{Number of iterations}

When measuring the self-diffusion, the system starts being empty. The smallest loading is measured first. We do $1. 10^7$ iterations (Monte Carlo steps) to equilibrate the system to the first loading. Then the measurement is started. This measurement consists of $2. 10^9$ iterations, for which the average flux and concentration difference of labeled particles between all the cavities is calculated.
For the second loading, which is a bit higher than the first, we start from the first loading, equilibrate $1. 10^7$ iterations and do a measurement of $2. 10^9$ iterations. This continues until the last loading. 

For the transport diffusion, the cavities at each loading are filled according to $p^{\mathrm{eq}}_n((\mu_l+\mu_r )/2)$, after which $2.10^6$ iterations are done to equilibrate. The measurements are also done for $2. 10^9$ iterations. 

This simulation is done 4 separate times. The first two and last two results are averaged. These two numbers give the error bars. The error bars are smaller than the symbols for all curves in the letter. Equilibration plus measurement takes around 30 minutes for one loading.

\section{$f(n)$ of Fig.\ $2(d)$.}

\begin{figure}
\centering
\includegraphics[width=0.6\textwidth]{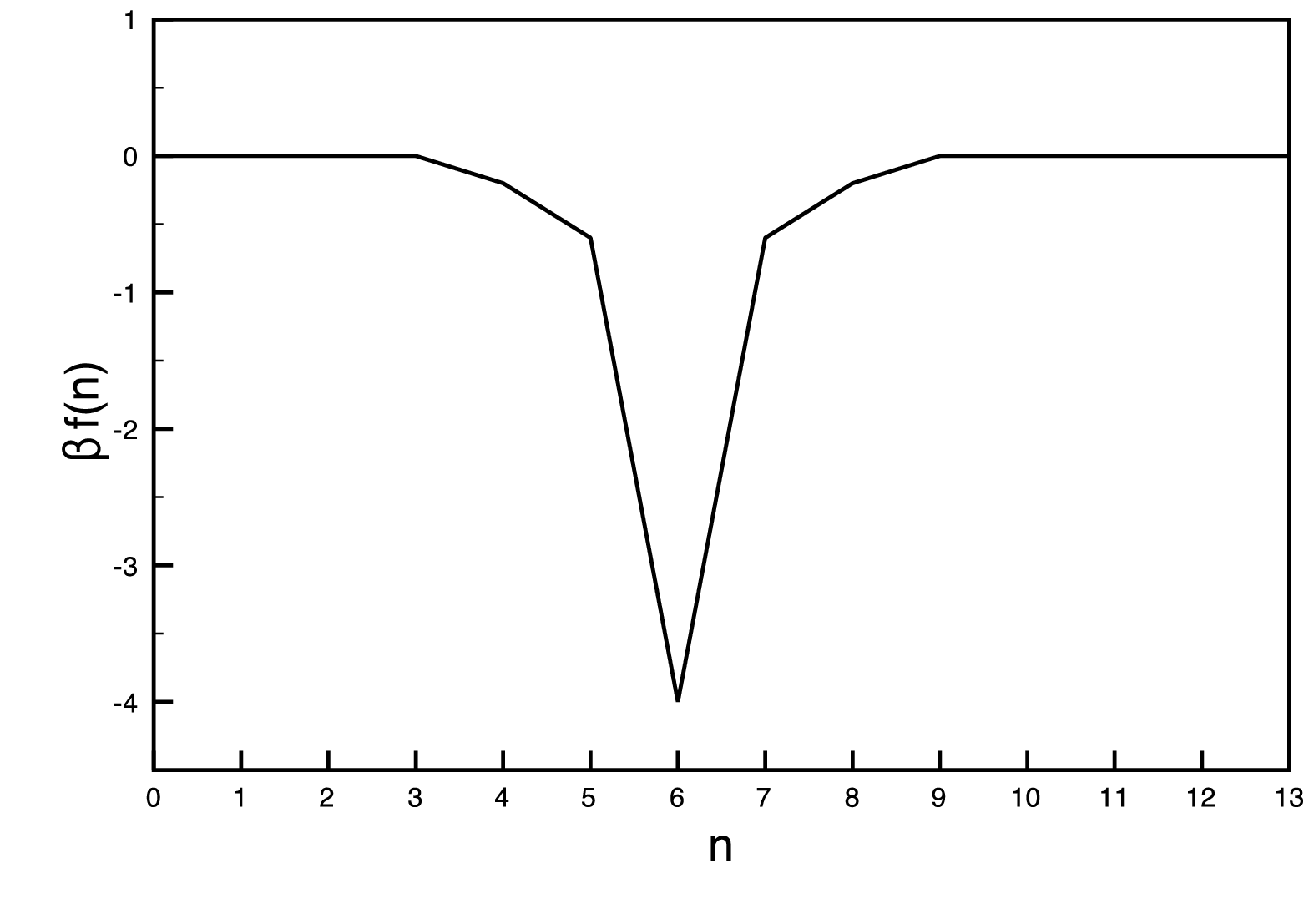}
\caption{$\beta f(n)$ of Fig.\ $2(d)$ in the letter.}
\label{fig::fn}
\end{figure}

The free energy of Fig.\ $2(d)$ is given in table \ref{tab::1}. A plot of this free energy is given in figure \ref{fig::fn}. From $0$ to $6$ $f(n)$ is concave, between $5$ and $7$ it is convex, and between $6$ and $13$ it is concave ($n_{max}=13$).

\begin{table}[h]
\begin{center}
\begin{tabular}{  l | c|c|c|c|c|c|c|c|c|c|c|c|c|c }
$n$ & 0 & 1 & 2 & 3 & 4 & 5 & 6 & 7 & 8 & 9 & 10 & 11 & 12 & 13 \\ \hline
$\beta f(n)$ & 0 & 0 & 0 & 0 & -0.2 & -0.6 & -4.0 & -0.6 & -0.2 & 0 & 0 & 0 & 0 & 0 \\
\end{tabular}
\caption{\label{tab::1} $\beta f(n)$ used in Fig.\ $2(d)$}
\end{center}
\vspace{-0.6cm}
\end{table}

\section{Fit with ethanol in ZIF-8}

\begin{figure}
\centering
\includegraphics[width=0.75\textwidth]{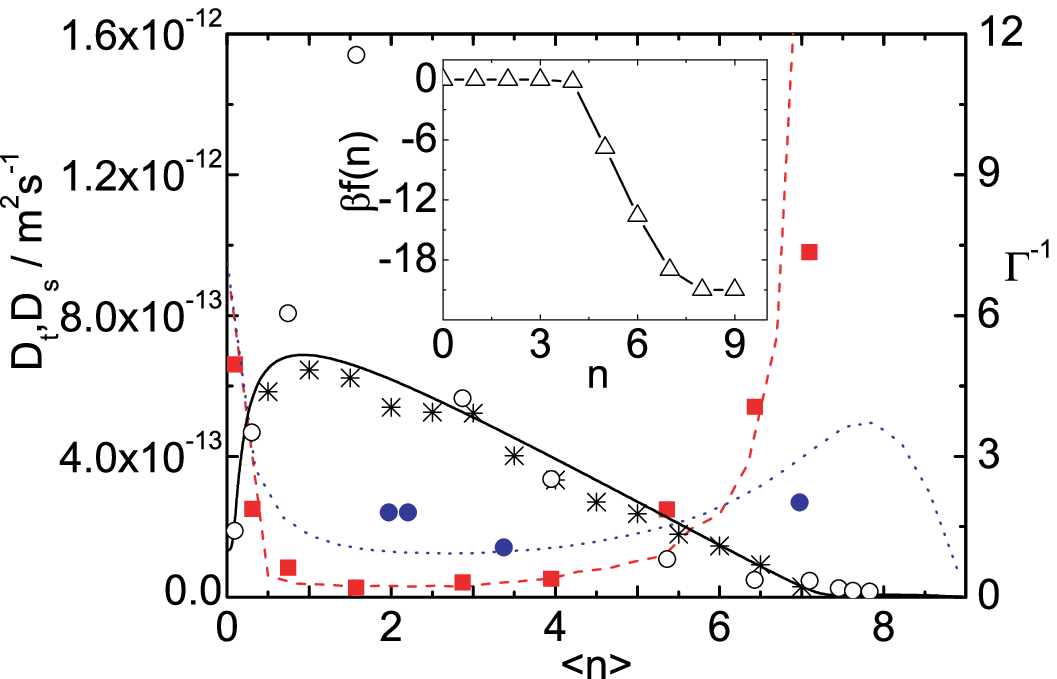}
\caption{Diffusion of ethanol in ZIF-8. The experimental self-diffusion is given by the blue full circles and the results from simulations by the blue dotted line. The experimental transport diffusion is given by the red squares and the results from simulations by the red dashed line (values on lhs axis).
The free energy function is shown in the inset. $\beta f(n)$ is represented by the black open triangles, the black line serves as a guide to the eye. The ratio between the self-diffusion and transport diffusion taken from the simulations is given by the black stars. The experimental $\Gamma^{-1}$ is given by black open circles, the analytical result is given by a full black line (values on rhs axis). Experimental data from \cite{communicationChmelik}.}
\label{fig::ethanol}
\end{figure}

To find a good free energy function for ethanol, we used a different fitting procedure than in the Letter. Using the analytical results for the thermodynamic factor and the diffusion coefficients as a guide, several free energies where tried until a good resemblance with experiment was found. Since calculation of the analytical results for different free energies requires no computation time, this resemblance can easily be checked 'by hand'.
$D_0$ is taken slightly higher than the experimental value of $D_t$ for the lowest measured loading ($D_0 = 9.6$ $10^{-13}$ $m^2/s$). $n_{max}$ is taken equal to 9. The values of $\beta f(n)$ are given in table \ref{tab::2}. The first part of $f(n)$ is constant, after which it becomes (very) concave. The last part is convex.
The simulation results and experimental data are given in figure \ref{fig::ethanol}. Good agreement is found with experiment, except for the outlier for $\Gamma^{-1}$ around $\langle n \rangle \approx 1.6$.

\begin{table}[h]
\begin{center}
\begin{tabular}{  l | c|c|c|c|c|c|c|c|c|c }
$n$ & 0 & 1 & 2 & 3 & 4 & 5 & 6 & 7 & 8 & 9 \\ \hline
$\beta f(n)$ & 0 & 0 & 0 & 0 & -0.2 & -6.8 & -13.6 & -19.0 & -21.0 & -21.0 \\
\end{tabular}
\caption{\label{tab::2} $\beta f(n)$ used in simulations for ethanol in ZIF-8.}
\end{center}
\vspace{-0.6cm}
\end{table}